\documentstyle[aps,epsfig,12pt]{revtex}
\begin{document}
\draft
\title{Entanglement and quantal coherence:
a study of two limiting cases of rapid system--bath interactions}
\author{Nicole F. Bell$^1$, R. F. Sawyer$^2$ and Raymond R. Volkas$^1$ }
\address{$^1$ School of Physics, Research Centre for High Energy Physics\\
The University of Melbourne, Victoria 3010 Australia\\
$^2$ Department of Physics, University of California at Santa Barbara \\
Santa Barbara, California 93106\\
(n.bell@physics.unimelb.edu.au, sawyer@vulcan.physics.ucsb.edu, r.volkas@physics.unimelb.edu.au)\\
Pacs 3.65w, 3.67a, 5.40a}
\maketitle

\begin{abstract}

We consider the dynamics of a system coupled to a thermal bath, going
beyond the standard two-level system through the addition of an energy
excitation degree of freedom. Further extensions are to systems containing
many fermions, with the master equations modified to take Fermi-Dirac
statistics into account, and to potentials with a time-dependent bias that
induce resonant avoided crossing transitions. The limit $Q \to \infty$,
where the interaction rate with the bath is much greater than all free
oscillation rates for the system, is interrogated. Two behaviors are
possible: freezing (quantum Zeno effect) or synchronization (motional
narrowing). We clarify the conditions that give rise to each possibility,
making an explicit connection with quantum measurement theory. 
We compare the evolution of quantal
coherence for the two cases as a function of $Q$, noting that full
coherence is restored as $Q \to \infty$. Using an extended master
equation, the effect of system-bath interactions on entanglement in
bipartite system states is computed. In particular, we show that the
sychronization case sees bipartite system entanglement fully preserved in
the large $Q$ limit.

\end{abstract}
\vskip2pc

\section{Introduction}

The dynamics of a quantum system coupled to a thermal bath is of great
interest in many branches of physics \cite{leggett}. This topic is of
practical importance because no system apart from the universe as a whole
is actually isolated. The contemporary focus on quantum computation
\cite{nielsen} has been providing added impetus to these studies, as has
the need to understand collision-affected neutrino oscillation dynamics in
astrophysical and cosmological settings \cite{reviews}. From a fundamental
perspective, the effect of system-bath interactions on quantal coherence
in the system bears on the measurement problem and on the emergence of
classical behavior out of an underlying quantum dynamics \cite{zurek}.

In this paper we will revisit two closely related phenomena that arise in
the limit of very rapid system-bath interactions: freezing (or quantum
Zeno effect) and motional narrowing (or synchronization). Our intentions
are twofold: (i) To present a unified perspective on these effects that
emphasizes their connection with each other, and their connections to
general quantal coherence and measurement questions.  (ii) To extend the
analysis to incorporate Fermi-Dirac statistics, a potentials with time
dependent bias, and 2-particle entangles states. For definiteness we will
focus on a double-well system, which turns out to be a versatile container
for exemplifying and coordinating a number of phenomena. However almost
all of our considerations can be carried over to other arenas, e. g. spin
systems.

\subsection{Freezing or Zeno}

What we call freezing in this paper, the drastic slowing of a process
through environmental interactions, can be discussed in a number of ways.
If we start with an oscillator of some kind, and crank up a damping
parameter to the point that the oscillator hardly budges in any reasonable
amount of time, we might call the system overdamped. If we start with a
decaying system on which we make very rapid measurements to see whether
the original state is still there and find that the system has hardly
evolved, we call it the quantum Zeno effect (QZE) \cite{zeno}. Except for
the fact that something has been slowed, two behaviors that are described
so differently appear to have little in common. But when the
``over-damped'' behavior in the oscillator comes from interactions with a
quantum thermal bath, and the ``observation" in the QZE is replaced by
interactions of the system with such a bath as well, then we find that the
two terminologies can describe the same essential physics. If we
particularize the ``oscillator'' in the above to the case of a particle in
a symmetrical double-well, then it turns out that we must require the bath
interaction to distinguish left from right in order for freezing to occur;
hereafter we shall refer to such a bath as ``side-sensing".
Correspondingly, for the particle-decay prototype of the QZE, in order to
simulate rapid measurements the bath must sense the presence or absence of
the initial particle.  For the widely discussed two-state systems, the
freezing phenomenon is essentially the same whether they are double-well
systems originating in a condensed matter context \cite{leggett},
molecules \cite{harrisandstodolsky} \cite{silbey}, nuclear spins, squids
\cite{silvestrini}, or neutrinos in the early universe
\cite{earlyuniverse}.
 
In practical circumstances we shall want to know the answer to
quantitative questions such as how frozen a system becomes as a result of
bath interactions, the answers to which can be roughly expressed in terms
of ratios of the different time scales in the problem. We also want to
know how the slowing or freezing behavior is correlated to other
calculable phenomena in the system.

\subsection{``Synchronization in a linear system'' or ``motional narrowing''}

These two effects seem to be identical at the formal level.  They carry
different names as a result of arising in very different physical
contexts. Synchronization is a natural partner to the slowing or freezing
phenomena for the case in which the bath does {\it not} distinguish which
of the two states the two-state system is in. We shall describe such a
coupling as ``side-blind''.  For a pure two state system, this describes
an interaction that does nothing at all of note. But in our example of an
electron in a symmetrical double well we may wish to place the system in a
bath that can excite or deexcite the states that are higher lying than the
first two (i.e. higher lying than the nearly degenerate symmetrical and
anti-symmetrical combinations of the single-well ground states). We shall
refer to the inclusion of these states as the introduction of a
``vertical'' coordinate, using ``horizontal'' to denote the left and right
coordinate. The oscillation rates are different for the higher-lying
states, so that the density matrix for a particle that is, say, initially
on one side of the well no longer oscillates sinusoidally between the two
states. But in the limit of large ``side-blind" coupling to the bath the
simple sinusoidal behavior is restored, as though the different
frequencies had been made the same, and the oscillations of the different
vertical components synchronized. Similar behavior in a neutrino-based
example has been reported in ref. \cite{bell}. ``Motional narrowing" has
been used to describe analogous behavior in an NMR system
\cite{motionnarrow}.
Qualitatively, motion narrowing occurs when the system is made to change state 
so rapidly that it is only able to respond to an average Hamiltonian.
That is, the ``horizontal'' evolution is governed by parameters which are averaged 
over the ``vertical'' states.

In this paper we review the basic analysis required to establish the above
assertions, including the derivation of a master equation, all in the
double-well context, and we provide some plots of typical behavior of the
solutions. Then we investigate a number of new topics related to the
limits mentioned above:

\subsubsection{Several fermions in the well} This entails writing a
modified master equation to take into account Fermi statistics. The
solutions, in the limit of large ``side-blind" noise, have a
counterintuitive property: an assemblage of several particles can move
sinusoidally from one side to the other, giving the appearance of being in
the same state.

\subsubsection{Systems with four horizontal states} This is like the
familiar case of two spin degrees of freedom; in our well realizations we
utilize two independent particles in the well, each particle with the same
structure of vertical levels.  We can now study entanglement of the
horizontal states. In the presence of moderate ``side-blind" noise we find
rapid destruction of entanglement, while in the ``synchronizing'' limit of
large ``side-blind" noise we have, in contrast, perfect preservation of
entanglement.  The latter may find application in quantum information
studies, just as its freezing cousin has already done \cite{refrigerator}.

\subsubsection{Avoided crossings} If we add a variable bias to the
symmetrical well, so that one side is deeper than the other, we can
perform interesting numerical experiments in which the bias is slowly
changed from one sign to the other. In the absence of the coupling to the
thermal bath, a particle localized on one side can be permanently
transported to the other side through an adiabatic reversal of sign of the
bias. When moderate ``side-blind" noise is added, we find that instead of
the near 100\% transportation of the uncoupled case we obtain a final
distribution with approximately 50\% of the probability on either side.
However, in the presence of strong ``side-blind" noise we find the
efficiency of the transportation to be restored.

\subsubsection{Quantum measurement} We discuss the connection between
system-bath interactions and quantum measurement. We show that the
system-apparatus-environment entanglement needed within the emerging
paradigm for the description of quantum measurements can arise naturally
in our context for the side-sensing case.

\subsection{Structure of the paper}
 
In section \ref{freezesynch} we specify our system-bath Hamiltonian and
the corresponding master equation for the system.  We then discuss both
the synchronized and frozen behavior and include an extension to the case
of multi-particle systems with Fermi statistics.  At the end of this
section we analyse the case of the double well with a time dependent bias,
focussing on avoided level crossing transitions.  In section
\ref{entangle} we determine the evolution of entangled well states under
the influence of side-blind noise.  We discuss the side-sensing and
side-blind interactions from a measurement theoretic perspective in
section \ref{meas} and in section \ref{conc} we conclude.

\section{Dynamics of freezing versus synchronization}
\label{freezesynch}

\subsection{The master equation: single particle case}

We focus on those situations where the effect of system-bath interactions on
the dynamics of the system can be derived from a master equation for the
reduced density matrix of the system.

The horizontal ``spin'' coordinate of the $n$-state system (where $n = 2$ for the double well)
is indexed with Greek letters, while the vertical energy variables will be denoted $E_i$.
The system basis states are denoted
$|E_i,\alpha \rangle$. The dynamics of the system uncoupled to the bath is governed
by the Hamiltonian,
\begin{equation}
H_0^{\rm sys} =
\sum _{i,\alpha,\beta} [E_i\delta_{\alpha \beta} + \lambda_{\alpha \beta}(E_i)]
|E_i,\alpha \rangle \langle E_i,\beta|.
\label{1}
\end{equation}
We will study the case $\lambda_{\alpha \beta}(E_i) \ll E_k$ for all values $i,k$.
This means that the energy eigenstates fall into nearly degenerate pairs near each $E_i$,
and the energy splittings
within these pairs are always much smaller than the spacings $\Delta E$ between the
pairs (single-well spacings).

Consider a reduced
density matrix for the system $\rho^{\rm sys}(t)={\rm Tr}_{\rm bath}[\rho^{\rm
s+b}(t)]$, where  $\rho^{\rm s+b}(t)$ is the complete density matrix of the model.
At an initial time  $t = 0$ we choose a form diagonal in the vertical
indices,
\begin{equation}
\rho^{\rm sys}(t=0)=\sum_{i,\alpha,\beta}  \rho_{\alpha\beta}(E_i,t=0)
|E_i,\alpha\rangle\langle E_i,\beta|.
\label{2}
\end{equation}
For these initial conditions, one of the results of a master equation derivation
is that for times $t \gg (\Delta E)^{-1}$,
the operator $\rho^{\rm sys}$ remains so nearly diagonal in the
vertical (energy) indices that we can continue to describe the system by a
vector in the energy space,  $\rho(E_i,t)_{\alpha\beta}$, replacing ($t=0$) by ($t$)
in Eq.(\ref{2}). Henceforth we suppress the indices $(\alpha,\beta)$ in the horizontal
space, in which $\rho(E_i,t)$ remains an operator.

We now have to supply a model for the system-bath interaction. A natural
and simple case to consider is
a factorised form,
\begin{equation}
H_{s-b}=\zeta V,
\end{equation}
where
$\zeta$ operates on the horizontal indices and $V$ depends only on the vertical
coordinates and the coordinates of the bath.
This coupling is also taken to be weak, in a way that allows
the problem to be solved, although it can at the same time be strong in another
sense: we will eventually study the limiting case where the system-bath
interaction rate is much larger than the free oscillation rates of the system.

In the Appendix, we show that for this form of coupling the master equations,
generalized to include the vertical structure, are of the form,
 \begin{eqnarray}
{\partial \over \partial t}\rho(E_i,t) & = & -i
[\lambda(E_i),\rho(E_i,t)]+\sum_{j}\zeta\, \rho(E_j,t)\,\zeta \,\Gamma(E_j,E_i)
\nonumber\\
& - & {1\over 2}\Bigr(\zeta^2\, \rho(E_i,t)+\rho(E_i,t) \,\zeta^2 \Bigr)
\sum_{j}\Gamma(E_i,E_j).
\label{3}
\end{eqnarray}
These equations are of Bloch form.
All of the elements in Eq.(\ref{3}) are matrices in the horizontal space except for the
rate functions $\Gamma$. They obey the relation
\begin{equation}
\Gamma(E_j,E_i) = \exp[(E_j-E_i)/T]\Gamma(E_i,E_j)
\end{equation}
due to the thermal equilibrium of the bath.
For the neutrino application, and with the right identifications,  Eq.(\ref{3}) gives
the ``quantum kinetic equations" derived by McKellar and Thomson \cite{MandT}.
They can also be found in the NMR literature, for example in Ref. \cite{motionnarrow}.
The derivation in the Appendix
produces the equations in the context needed for our present results.

We now specialize to the symmetrical double well. The two-dimensional horizontal space
consists of the left and right sides of the well, conveniently labelled by the
eigenvalues of $\sigma_3$, namely $+1\ (-1)$  for the particle to be on
the left (right) side. The barrier height is chosen to be large
compared to $\Delta E$.
The vertical states are the single well energy excitations.  We take
\begin{equation}
\lambda_{\alpha\beta}(E_i)=[\sigma_1]_{\alpha\beta}g(E_i)
\end{equation}
in order to model the free oscillations between left and right.
The energy splittings $g(E_i)$, which are essentially tunneling rates, have strong
$E_i$ dependence.
For the ``spin'' dependence of the coupling to the bath we take,
\begin{equation}
\zeta = 1 + b \sigma_3,
\end{equation}
where the parameter $b$ plays the important role of modelling the
horizontal structure of the system-bath interaction.
In the case $b = 0$, the transition matrix
elements of
the operator $H_{\rm s-b}$
for $E_i\rightarrow E_j$ in the left-hand well, considered by
themselves, are the same as those for the right-hand well, as would be the case in the
dipole approximation of the interaction with a radiation field. For the case $b = 1\ (-1)$,
only the amplitude on the left-hand (right-hand) side of the well interacts with the bath. In
other words, $b = 0\ (\pm 1)$ describes left-right symmetric (maximally asymmetric)
system-bath couplings: side-blind (maximally side-sensing) in other words. 
As we review below, frozen behavior arises as a limiting case
for $b \neq 0$ couplings, while synchronization or motional narrowing
occurs as a limiting case when $b = 0$.

The function $\Gamma(E_i,E_j)$ is the (single well) rate
for a state $E_i$ to make a transition to state $E_j$.
When $T \approx \Delta E$, the bath induced transitions among the vertical
states are dynamically important. In this case our
models differ substantively from models that do
not have the vertical structure.


As an example we use a double square well with infinite barriers to the left and right
of $x = \pm 8a$, and with a
central barrier of width $2a$ and height $U_0$. We choose  $U_0$ and the particle
mass such that
twenty states are bound with energies less than $U_0$. The
low-lying states are thus very nearly the symmetrical and anti-symmetrical
combinations of
separate states in an infinite well of width $7a$. The energy splittings, $g(E_i)$
are easy to calculate
in the limit of small tunneling.

For the bath we take bosons with $\omega=vk$,
equal energy spacing
 $2\pi v/L$ where $L\rightarrow \infty$, and with creation and annihilation
operators, $a_n ^\dagger ,
a_n$. We take
\begin{equation}
H_0 = \sum_n \omega_n a_n ^\dagger a_n  + H_0^{({\rm sys})},
\end{equation}
and in the
system-bath interaction, $H_{s-b}=V\zeta$, we take the vertical operator $V$ to
have off-diagonal matrix elements given in dipole expansion as
\begin{equation}
V_{jk}=
 {q\over L}\sum_n[(a_n ^\dagger +a_n )\delta_{jk}+ik_nx_{jk}(a_n^\dagger - a_n )].
\label{4}
\end{equation}
where $q$ is a coupling strength.
We calculate the off-diagonal dipole matrix elements, $x_{jk}$, using the infinite
well wave-functions and define $c_{jk}=q k_n x_{jk}$,
where the later formalism will justify using (near) energy conservation to
express $k_n$ as $\pm(E_j-E_k)/v$. The transition rates in the bath interaction part
of Eq.(\ref{3}) are then,
\begin{equation}
\Gamma(E_k,E_j) =2 \pi c_{jk}c_{kj}\Bigr[{ \theta(E_j-E_k)
\over e^{(E_j-E_k)/T}-1}+ {\theta(E_k-E_j) \over 1-e^{(E_j-E_k)/T}}\Bigr].
\label{11}
\end{equation}

We will now display solutions to the master equations for the cases $b = 0$ and $b \neq 0$.

\subsection{Results for $b=0$: synchronization or motional narrowing}

We first consider side-blind system-bath couplings, so that $b = 0$, and
we solve Eq.(\ref{3}) for several values of $q^2$.  The
resulting curves are labeled
with a ratio $Q$ that is proportional to $q^2$,
\begin{equation}
Q \equiv \langle \Gamma \rangle / \langle g \rangle,
\end{equation}
where $\langle g \rangle$ is the thermal
average of the oscillation rates, $g(E)$, and $\langle \Gamma \rangle$ that of the bath induced
transition rates, $\Gamma(E_j,E_k)$. Figure 1 shows the probability that a particle
beginning on the left side in a thermal distribution will be on the left side at time $t$. We choose
temperature $T = 5$ in units of the single-well ground state energy.

\begin {figure}[h]
    \begin{center}
        \epsfxsize 3in
        \begin{tabular}{rc}
            \vbox{\hbox{
$\displaystyle{ \, { } }$
               \hskip -0.1in \null} 
} &
            \epsfbox{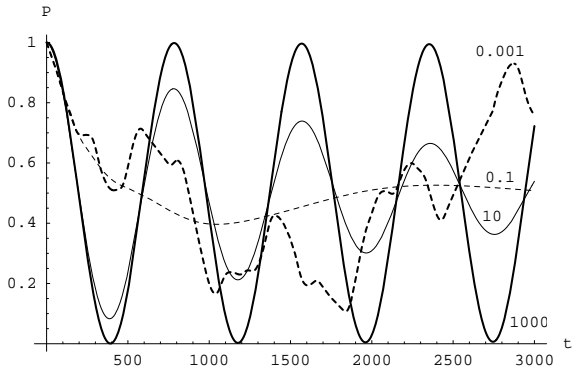} \\
            &
            \hbox{} \\
        \end{tabular}
    \end{center}
\label{fig1}
\protect\caption
	{%
The probability $P$ for a particle to be found in the left hand well,
for a bath temperature $T=5$, in units of the ground state energy of the single infinite well.
The initial condition is that the particle is
on the left with a thermal distribution of energies. The curves are labelled by
values of the rate-ratio parameter $Q$ that range from $0.001$ to $1000$.
The time is in an arbitrary unit, the scale of which is
of the order of $\hbar$ times the inverse of the single well energy 
spacings.
	}
\end {figure}

For the largest value of $Q$ the probability shows undamped sinusoidal behavior in time. (This can be
demonstrated in the $Q \rightarrow \infty $ limit analytically, with the result that the oscillation
frequency is the thermal average of the oscillation frequencies for the separate modes \cite{bell}).
For somewhat smaller values the probability settles rather quickly to 50\%. For very small $Q$ the
probability shows the irregular behavior characteristic of the incommensurate frequencies $g(E_j)$.
Note that arbitrarily large values of $Q$ can be achieved for any value of system-bath coupling, $q$
(whether the mechanism be photons, phonons or collisions with surrounding particles), by widening
sufficiently the barrier between the wells.

In Fig.\ 2 we show the von Neumann entropy, as defined by
\begin{equation}
S = - \sum_j{\rm Tr} \{\rho(E_j)\log[\rho(E_j)]\},
\end{equation}
where the
trace is over the two-dimensional horizontal space.
We note that as $Q$ is increased from small to moderate values the
rate of entropy increase rises, but that as $Q$ becomes very large the rate
of entropy increase goes to zero.

\begin {figure}[h]
    \begin{center}
        \epsfxsize 3in
        \begin{tabular}{rc}
            \vbox{\hbox{
$\displaystyle{ \, {  } } $
               \hskip -0.1in \null} 
} &
            \epsfbox{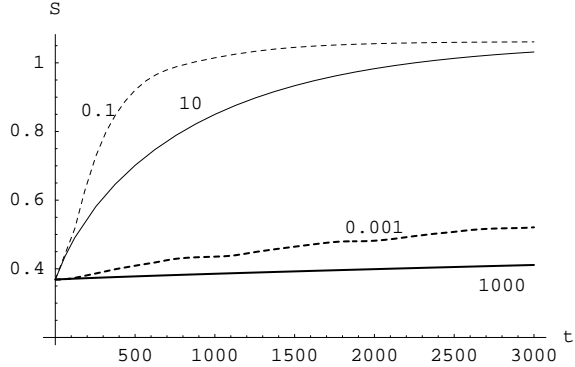} \\
            &
            \hbox{ } \\
        \end{tabular}
    \end{center}
\label{fig2}
\protect\caption
	{%
The entropy S as a function of time for the same values of $Q$ as used in fig.1.
Time in arbitrary units.
	}
\end{figure}

Notice that the entropy begins at a nonzero value. This is simply due to
our initial state being thermal with respect to the vertical variable.
(If we had chosen a non-thermal state it would have been rapidly thermalized
anyway in the large $Q$ limit.) Any increase in $S$ beyond this base level
then reflects loss of order with respect to the horizontal variable, which
is really what we are interested in.

\subsection{Results for $b\ne 0$: freezing or quantum Zeno effect}

We briefly summarize some examples. In the case plotted in Fig.\ 1, for the strongest coupling, we find
little change for $b = 0.001$.  But $b = 0.005$ starts to poison
the oscillation, damping it by a factor of
20\% in five periods.  As we move to much larger $b$, the motion becomes more frozen, so that when,
for example,
$b = 0.5$, 98\% of the probability has remained on the left hand side over a time span of five average
periods of the $b = 0$ system. Thus side-sensing bath interactions  are very
efficient in freezing the time development.

The von Neumann entropy $S$ shows qualitatively similar behavior to the $b = 0$ case.
As we increase $R$, the entropy at first increases, but eventually it becomes frozen at its
initial value when the QZE limit is reached. An important difference between the two
cases is that entropy increase abates for $b = 0$ with the system evolving non-trivially
with time, whereas for $b \neq 0$ it freezes, simply because all evolution freezes.

\subsection{Fermi statistics case}

We presented the results for the previous sections in terms of the density matrix
for a single particle in the well. If we choose to put several distinguisable particles 
in the same well and there are no interactions among these particles, then the results 
for each particle are the same as the previous results, and if we wished we could write a 
single equation for the density matrix for the assemblage, normalized to the number of particles.
The same is true for the case of an assemblage of indistinguishable particles in the case that
the temperature is high enough for Boltzmann statistics to obtain. For the $N$ particle case
with Fermi statistics, in the regime where the statistics matter, we have not succeeded in deriving
an acceptable modification of  Eq.(\ref{3}). However, if we choose to use a formulation in which the diagonal matrix element of a density operator gives the probability that a state is occupied,  and these occupancies are taken as independent \footnote{When the occupancies are taken as independent, the particle number is necessarily indeterminate.}, there is a simple extension of  Eq.(\ref{3}) that describes the time evolution of the ensemble of fermions in the well. 
Following the argument
of \cite{raffelt} we replace Eq.(\ref{3}) by, 
 \begin{eqnarray}
 &{\partial \over \partial t}\rho(E_i,t)  =  -i
[\lambda(E_i),\rho(E_i,t)]+
\nonumber\\
&{1\over 2}\sum_{j}\Bigr( [1-\rho(E_i,t)]  \zeta\, \rho(E_j,t)\,\zeta \,+ \zeta\, \rho(E_j,t)\,\zeta \,[1-\rho(E_i,t)] 
 \Bigr )\Gamma(E_j,E_i)
\nonumber\\
& -  {1\over 2}\sum_{j}\Bigr(\zeta [1-\rho(E_j ,t)] \zeta\, \rho(E_i,t)+\rho(E_i,t) \,\zeta[1-\rho(E_j ,t)]  \zeta \Bigr)
\Gamma(E_i,E_j).
\label{masterfermi}
\end{eqnarray}
The density matrix is now to be normalized to the average number of particles, $\sum_j {\rm Tr}[\rho (E_j,t)]=\langle N\rangle$,
which is time independent.  As an example we take a system which begins at $t=0$ with no particles on the RHS, and an average number of particles $\langle N \rangle = 8$ on the LHS, for a well system with the same parameters used to generate data for the single particle calculation. For the initial energy distribution we take 50\% occupancy for each of the lowest sixteen states.

In the case of zero coupling to the thermal bath, the various particles rattle back and forth independently 
with their incommensurate frequencies. Our terminology has been changed here, since previously 
we pictured the density matrix for a single particle performing exactly the same gyrations. We note that 
in the case without vertical transitions the Fermi statistics enter only in the restrictions put on the initial state.

\begin {figure}[h]
    \begin{center}
        \epsfxsize 3.5in
        \begin{tabular}{rc}
            \vbox{\hbox{
$\displaystyle{ \, {  } } $
               \hskip -0.1in \null} 
} &
            \epsfbox{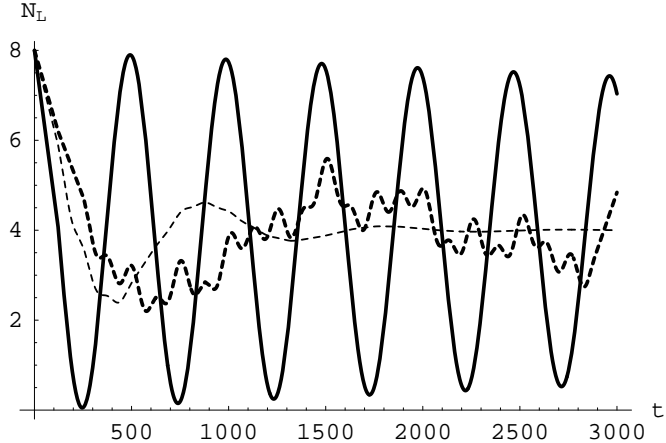} \\
            &
            \hbox{ } \\
        \end{tabular}
    \end{center}
\protect\caption{\label{fermi}
The time development of the average number of particles in the left hand well.
The initial conditions are: the first sixteen levels of the left well each occupied with
50\% probability. The heavy dashed curve gives the behavior
with extremely small coupling to the bath,  the lighter dashed curve the behavior
with an intermediate value of the coupling, and the heavy solid curve the
behavior with the largest coupling. The respective couplings are in the ratios
of $10^{-3}$, $1$, and  $10^{3}$.  The time is in arbitrary units.}
\end{figure}

In the presence of a moderate coupling to the bath, the system tends rather 
quickly to an equilibrium state with 50\%-50\% L-R occupancy, as shown in the 
heavy dashed curve in Fig.\ \ref{fermi}.
In this state the final energy distribution is a Fermi function,
\begin{equation}
\rho(E_j,t)= [1+e^{(E_j -\mu )/T}]^{-1}
\end{equation}
where the parameter $\mu $ is set by
\begin{equation}
2\sum_j [1+e^{(E_j -\mu )/T}]^{-1}= \langle N \rangle
\end{equation}

But, as before, if we increase the bath coupling sufficiently, the particle density oscillates 
back and forth sinusoidally as shown in the heavy solid curve of Fig.\ \ref{fermi}. 
The average frequency is now given by the average of the 
individual-state rates with respect to the Fermi distrubution function.  
We have not found in the literature an analogue of this behavior, 
in which a group of fermions undergoes collective synchronized oscillations, as though they 
were all in the same state.

\subsection{Adiabatic inversion}

A generalization of the above considerations is
the extension to a double well with a time dependent bias,
\begin{equation}
\delta H_0^{\rm sys}= \epsilon(t) \sigma_3.
\end{equation}
Consider a case in which we begin
with $\epsilon>0$ and $\epsilon \gg| g (E_i)|$ for some set of low-lying states 
(but $\epsilon\ll \Delta E$). A particle not interacting with the medium, initially 
localized on the left, with its density matrix distributed among the low-lying
states, will stay largely on the left as long as the bias is maintained at the initial 
value. If the bias is slowly lowered, going through zero and ending at the negative of 
the original value, then the particle will efficiently be carried to the right hand well. 
The mechanism can be described 
as ``avoided crossing''.  \footnote{This is the analogue of the
Mikheyev-Smirnov-Wolfenstein (MSW)  \cite{msw} transition in neutrino physics.
This behavior is addressed in \cite{silvestrini} for the case of a 
thermal bath that does distinguish the right and left sides of the well. As 
expected, their results are quite different for the case of large $Q$.} 
Now we consider the same process in the presence of a side-blind thermal
bath. With a moderate coupling to the bath, the system will still remain localized on the 
left if we keep the bias constant at its initial large value. But if we proceed as before
 through an adiabatic reversal of the bias, then the jostling caused by the vertical transitions 
will reduce the efficiency of the left to right
transportation, and will tend to leave one-half of the probability behind in the left hand well.

\begin {figure}[h]
    \begin{center}
        \epsfxsize 3.5in
        \begin{tabular}{rc}
            \vbox{\hbox{
$\displaystyle{ \, {  } } $
               \hskip -0.1in \null} 
} &
            \epsfbox{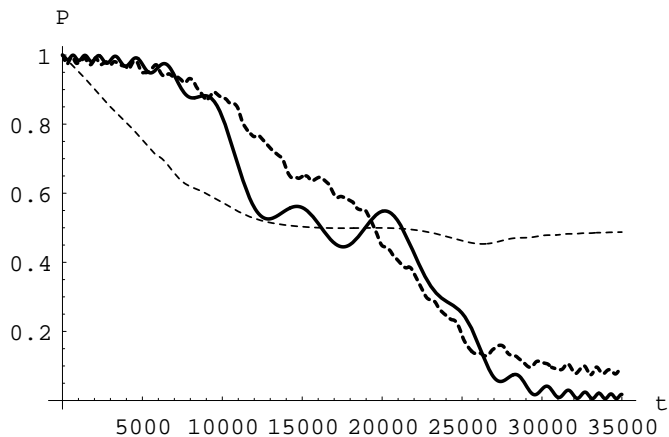} \\
            &
            \hbox{ } \\
        \end{tabular}
    \end{center}
\protect\caption{\label{bias}
The probability that the particle is on the left,  where the well bias is reversed. 
The heavy dashed curve is for zero system-bath coupling, the light dashed curve for 
moderate coupling and the solid curve for large coupling.  The time is in arbitrary
units.
(The dependence of $\epsilon$ on $t$ has been taken to be proportional to $(t-t_0)^3$, 
where $t_0$ is the zero-bias time.) 
}
\end{figure}

However, if we increase the strength  of the bath coupling sufficiently, then we regain the
efficient shift from left to right.\footnote{Detailed analysis of similar behavior for a neutrino 
application is given in \cite{bell}.}
Figure \ref{bias} shows the results of  the calculation, for the cases of no bath coupling,
moderate bath coupling and strong bath coupling. Note that in this example the transport of 
the particle from the left to right is even more efficient in the case of large bath coupling 
than it was in the case of no bath coupling, the small inefficiency
caused by some non-adiabaticity 
in the bias-changing process being reduced in the presence of the bath coupling.

This rather surprising behavior can perhaps be understood qualitatively in the following way: First consider the previously discussed synchronized oscillation limit for the symmetrical well. There the synchronized frequency is somewhat higher than we would have guessed from the oscillation frequency for a typical particle (with $E\approx T$ ) in the absence of of the bath, because in our
model the tunnelling rates increase rapidly as we move up in energy.  It is also true that the non-adiabaticity, as defined by the fraction of particles ending up on the left at the end of our bias-reversal, declines rapidly with increasing energy, again in the absence of bath interactions. Thus the bath's effects, still a bit mysterious, are to use the synchronized frequency for the purpose
of the estimation of adiabaticity. We emphasize that the real energy distribution of the particles remains very nearly thermal; but synchronization lets us do the tunnelling, in effect, at a higher energy, with consequent smaller inefficiency.

This phenomenon offers some promise for obtaining greater L-R transition efficiency
in an application. Another example of parameters,
beyond the ones presented in Fig.\ \ref{bias} is the following: take the temperature so low that the occupancy of the ground state 
is 97\%, the first excited state, 3\% and the higher states negligible. Then we find, for a particular rate of bias inversion and a particular level of coupling to the bath, that there is  6\% probability to end in the left hand well. This is to be compared to
16\% in the absence of the bath coupling.

\section{Evolution of entanglement}
\label{entangle}

Irrespective of the value of $b$, we have concluded that the quantum coherence
of a state that is initially a left or right eigenstate (with a thermal distribution of
energies) is preserved  under time evolution in
the $Q \to \infty$ limit.  For $b \neq 0$, this is because of freezing, while
for $b = 0$ coherence manifests as motional narrowing or synchronization.

It is natural to then ask whether entanglement is preserved in the same
limit.  For the QZE case, the answer is no, simply because any L-R
entangled state has off-diagonal entries that are rapidly damped when $b
\neq 0 $.  In other words, entanglement will be destroyed, not frozen, in
the QZE limit.  We now want to see what happens in the synchronization
case.  Intuitively, one expects preservation to be maintained because all
components of the entangled multi-particle system will oscillate between
the wells at the same rate.\footnote{Note that synchronization preserves
coherence given any initial state, as opposed to the QZE which preserves
coherence only if the initial state is an eigenstate of $\sigma_3$.} This
intuition is borne out by direct computation.

We will focus on bipartite or two-particle states for simplicity. 
The maximally entangled or Bell states are
\begin{eqnarray}
|\Psi^{\pm}, E_i, E_j  \rangle & \equiv & { |E_i,\, L \rangle \otimes |E_j,\, R \rangle \pm
|E_i,\, R\rangle \otimes |E_j,\, L\rangle \over \sqrt{2} },\nonumber\\
|\Phi^{\pm}, E_i, E_j \rangle & \equiv & { |E_i,\, L \rangle \otimes |E_j,\, L \rangle \pm
|E_i,\, R\rangle \otimes |E_j,\, R\rangle \over \sqrt{2} },
\label{bellstates}
\end{eqnarray}
where one well particle has energy $E_i$ and the other energy $E_j$.

A straightforward generalisation of the derivation sketched in the
Appendix yields the master equation for a general two-particle system.
We denote the basis states of the system as $| E_i,E_j',\alpha,\beta \rangle$
and consider a density matrix (diagonal in the two energy spaces)
$ \rho(E_i,E'_j)_{\alpha,\alpha ',\beta,\beta '}$.

Suppressing the flavor (R-L) matrix indices, the  Bloch equations,
generalized to include the vertical structure, are of the form,
 \begin{eqnarray}
{d\over dt}\rho(E_i,E_j,t)&=&-i [\lambda(E_i),\rho(E_i,E_j,t)]-i
[\lambda'(E_j),\rho(E_i,E_j,t)]
\nonumber\\
&+&\sum_{E_k}\zeta\,
\rho(E_k,E_j,t)\,\zeta \,\Gamma(E_k,E_i)+\sum_{E_m}\zeta'\,
\rho(E_i,E_m,t)\,\zeta '\,\Gamma(E_m,E_j)
\nonumber\\
&-&{1\over 2}\Bigr(\zeta^2\, \rho(E_i,E_j,t)+\rho(E_i,E_j,t) \,\zeta^2 \Bigr)
\sum_{E_j}\Gamma(E_i,E_k)
\nonumber\\
&-&{1\over 2}\Bigr((\zeta')^2\, \rho(E_i,E_j,t)+\rho(E_i,E_j,t)
\,(\zeta')^2 \Bigr)
\sum_{E_m}\Gamma(E_j,E_m).
\end{eqnarray}

Here $\lambda (E)$, $\lambda '(E)$ , $\zeta$, and $\zeta '$ are matrices in the (R-L, R-L) space.
For the case of symmetric double wells, the only horizontal matrices in the free oscillation part of
the Hamiltonian are constructed from the $\sigma_1$ matrices of the two particles and we have
\begin{eqnarray}
\lambda_{\alpha \alpha ' \beta \beta '}(E)=g(E)[\sigma_1]_{\alpha \alpha'}\delta_{\beta \beta'},
\nonumber\\
\lambda'_{\alpha \alpha ' \beta \beta '}(E')=g(E')\delta_{\alpha \alpha'}[\sigma_1]_{\beta \beta'}.
\end{eqnarray}

It is useful to parametrise the the density matrix in terms of the coefficients $x_{ij}$ in
an expansion in terms of sigma matrices
\begin{equation}
\rho = x_{ij}(E,E') \sigma_i \otimes \sigma_j.
\end{equation}
If we choose as our initial condition one of the four Bell states in eq.(\ref{bellstates}), we need
consider only the six components $x_{00},x_{11},x_{22},x_{23},x_{32}$ and $x_{33}$. If,
additionally,  the initial state is one of thermal equilibrium with respect to the vertical variables
then $x_{00}(E,E')$ and $x_{11}(E,E')$ remain constant and we are left with the four equations
\begin{eqnarray}
\frac{d}{dt} x_{ij}(E,E') &=& g(E)\sum_k \xi_{ik} x_{kj}(E,E') + g(E')\sum_k \xi_{jk} x_{ik}(E,E')
\nonumber \\
&+& \sum_{E_k} \left[ x_{i,j}(E_k,E_j',t) \Gamma(E_k,E_i)
- x_{ij}(E_i,E_j',t) \Gamma(E_i,E_k) \right]
\nonumber \\
&+& \sum_{E_m'} \left[ x_{ij}(E_i,E_m',t)\Gamma(E_m',E_j')
-x_{ij}(E_i,E_j',t) \Gamma(E_j',E_m') \right],
\label{exp}
\end{eqnarray}
where the nonvanishing elements of the matrices $\xi$ are $\xi_{32}=-\xi_{23}=2$.
In eq.(\ref{exp}) we have specialised the collision terms to the case $\zeta=\zeta'=1$,
in which the system-bath couplings do not distingiush the value of the (L,R) index.

We quantify the degree of entanglement using the {\it entanglement of formation} \cite{wootters}.
For a bipartite system, the entanglement of formation is given by
\begin{equation}
E_f(C)=h(\frac{1}{2}+ \frac{1}{2} \sqrt{1-C^2}),
\end{equation}
where
\begin{equation}
h(x)=-x\log_2x -(1-x)\log_2(1-x).
\end{equation}
The ``concurrence'' $C$ is
\begin{equation}
C = \text{max} \{0, \lambda_1-\lambda_2-\lambda_3-\lambda_4\},
\end{equation}
where the $\lambda_i$ are the square roots of the eigenvalues of the matrix $\rho \tilde\rho$,
and the spin-flipped state $\tilde \rho$ is defined as
\begin{equation}
\tilde\rho = (\sigma_y \otimes \sigma_y) \rho^* (\sigma_y \otimes \sigma_y).
\end{equation}

We illustrate the evolution of the entanglement in the figures below,
for the side-blind case with 
the initial conditions $|\Phi^{\pm}(E_1,E_1)\rangle$.  In the
absence of
the heat bath, a system initially in the state $|\Phi^{-}(E_1,E_1)\rangle$
will remain in that state. An initial state of
$|\Phi^{+}(E_1,E_1)\rangle$, however, evolves as $|\psi (t) \rangle =
\cos[2g(E_1)t]|\Phi^+(E_1,E_1)\rangle - i\sin[2g(E_1)t]|\Psi^+(E_1,E_1)\rangle $ 
in the absence
of a heat bath, while, of course, maintaining maximal entanglement.

\begin {figure}[h]
    \begin{center}
        \epsfxsize 3.5in
        \begin{tabular}{rc}
            \vbox{\hbox{
$\displaystyle{ \, {  } } $
               \hskip -0.1in \null} 
} &
            \epsfbox{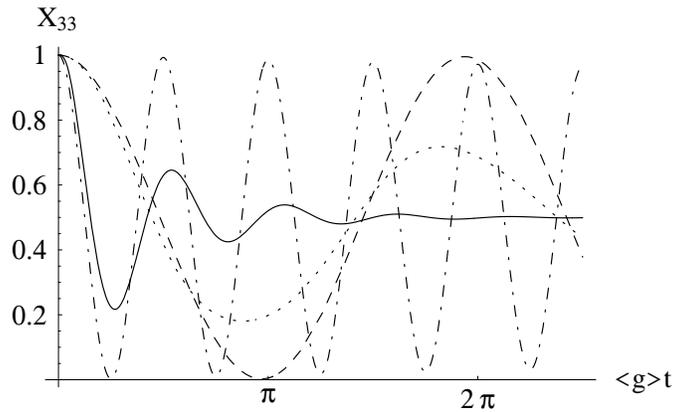} \\
            &
            \hbox{ } \\
        \end{tabular}
    \end{center}
\protect\caption{\label{x33s}
The evolution of $X_{33}$ as a function of time, for the initial state
$|\Phi^{+}(E_1,E_1)\rangle$.  The dashed, dotted, solid and dot-dashed
curves are for
$Q=$ 0.0005, 0.05, 5 and 500 respectively.}
\end{figure}

\begin {figure}[h]
    \begin{center}
        \epsfxsize 3.5in
        \begin{tabular}{rc}
            \vbox{\hbox{
$\displaystyle{ \, {  } } $
               \hskip -0.1in \null} 
} &
            \epsfbox{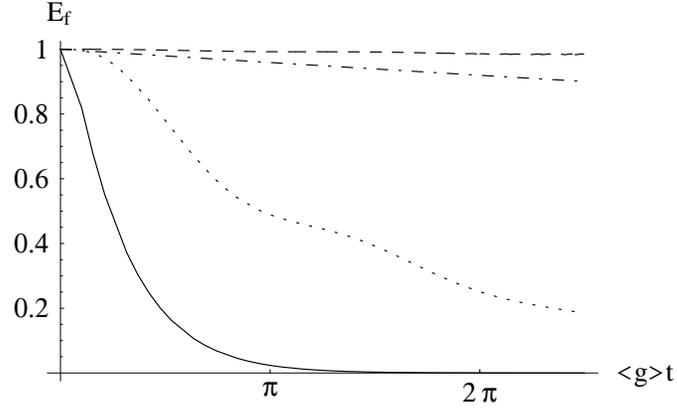} \\
            &
            \hbox{ } \\
        \end{tabular}
    \end{center}
\protect\caption{\label{es}
The entanglement of formation, $E_f$, for the same parameters as Fig.\ref{x33s}.}
\end{figure}

\begin {figure}[h]
    \begin{center}
        \epsfxsize 3.5in
        \begin{tabular}{rc}
            \vbox{\hbox{
$\displaystyle{ \, {  } } $
               \hskip -0.1in \null} 
} &
            \epsfbox{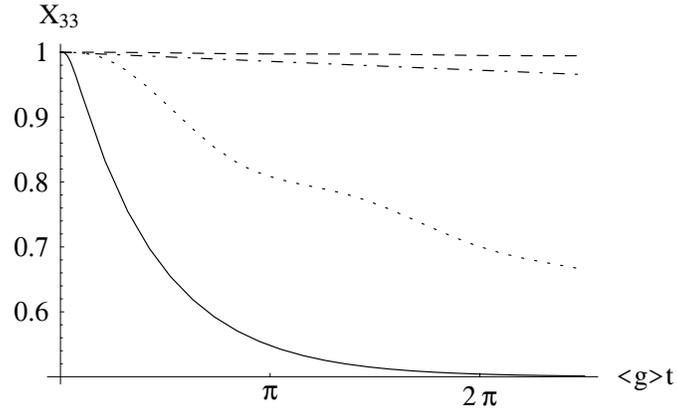} \\
            &
            \hbox{ } \\
        \end{tabular}
    \end{center}
\protect\caption{\label{x33a}
The evolution of $X_{33}$ as a function of time, for the initial state
$|\Phi^{-}(E_1,E_1)\rangle$. The values of the parameter $Q$ are as in Fig.\ref{x33s}.}
\end{figure}

\begin {figure}[h]
    \begin{center}
        \epsfxsize 3.5in
        \begin{tabular}{rc}
            \vbox{\hbox{
$\displaystyle{ \, {  } } $
               \hskip -0.1in \null} 
} &
            \epsfbox{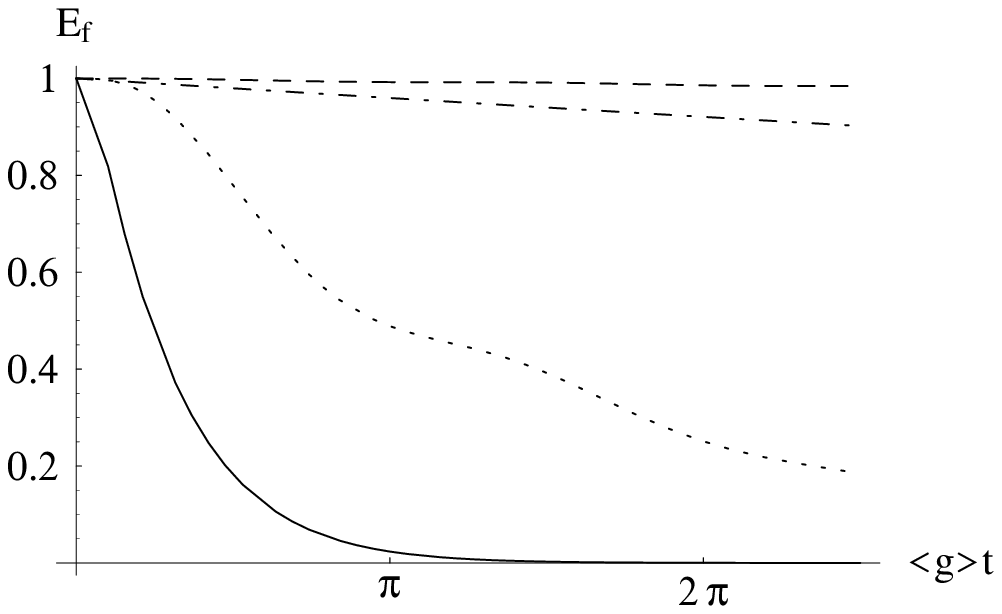} \\
            &
            \hbox{ } \\
        \end{tabular}
    \end{center}
\protect\caption{\label{ea}
The entanglement of formation, $E_f$, for the same parameters as in Fig.\ref{x33a}.}
\end{figure}

In Figs.\ \ref{x33s} and\ref{x33a} we plot the evolution of $X_{33}$,
\begin{equation}
X_{33}= 2\left(0.25+\sum_{i,j}x_{33}(E_i,E_j) \right),
\end{equation}
which measures the probability that if one of the particles is on the left (right), the other
particle will also be found on the left (right).
In Figs.\ \ref{es} and \ref{ea} we plot the entanglement of formation, calculated
after first summing over
the vertical states
\begin{equation}
\rho = \sum_{i,j} \rho(E_i,E_j).
\end{equation}

For small to moderate $Q$ values, we observe the gradual relaxation and
decay of entanglement arising from the population of energy levels with
different frequencies. For large values of $Q$, we obtain synchronised
behaviour, where the evolution of the system resembles that of two
isolated qubits, with $\lambda(E_1)$ replaced by an averaged frequency
$\overline{\lambda}$. That is, despite the strong coupling to the bath,
the system behaves, effectively, as though it were isolated.
In the limit $Q \rightarrow \infty$, we see the persistance of maximal
entanglement.

Finally, note that entanglement may be created or destroyed if the
Hamiltonian were to have a term describing an interaction between the two
qubits.  In the large $Q$ limit this process will also proceed as if the
system were isolated, with the Hamiltonian parameters replaced by their
averaged values.

\section{Discussion}
\label{meas}

The freezing phenomenon or QZE is well known. It should be noted, though,
that much of the recent non-neutrino literature does not
study systems with a vertical structure. Indeed, when the
vertical structure is removed Eq.(\ref{3}) reduces to the familiar form,
\begin{equation}
\frac{\partial \rho}{\partial t} = -i [\, \lambda,\, \rho\, ] -
\frac{\Gamma}{2}[\, \zeta,\, [\, \zeta,\, \rho\, ]\, ].
\end{equation}
The decoherence term is then simply
\begin{equation}
\frac{\Gamma}{2}[\, \zeta,\, [\, \zeta,\, \rho\, ]\, ] = - 2 b^2 \Gamma
\left( \begin{array}{cc} 0 & \rho_{12} \\ \rho^{*}_{12} & 0
\end{array} \right),
\end{equation}
from which freezing can
be deduced in the large $Q$ limit, provided that $b \neq 0$.
The presence of vertical states evidently
does not alter these qualitative conclusions:
freezing occurs at large $Q$ for $b \neq 0$ whether vertical structure
exists or not.

The above derivation also shows that $b = 0$ system-bath interactions have no affect
on the system in the absence of vertical structure. But, when it is not
absent, we have seen that there is an interesting conceptual
cousin to the $b \neq 0$ quantum Zeno effect: 
synchronization (or motional
narrowing) when $b = 0$.\footnote{Of course this manifests itself only when the vertical
coordinate exists because one cannot talk about synchronization for a one-oscillator
system!} Both occur in the $Q \to \infty$ limit, when the dynamics of the
system is dominated by its interactions with the bath.
While synchronization from
motional narrowing is a known phenomenon, we have not found a clear and
comprehensive discussion in the literature about its relationship with the
QZE, quantal coherence and measurement, so we now provide one.

The strong connection between freezing/synchronization and coherence is revealed
by the evolution of entropy. Pure states always have vanishing entropy.
Decoherence in the system induced by system-bath interactions can cause initially pure
states to evolve into mixed states. In this case the entropy increases towards
a maximum which is attained when all states in the statistical mixture
are equally populated, with complete decoherence.

We have to be careful to distinguish between what one may call vertical
and horizontal contributions to entropy. As explained when we discussed Fig.\ 2,
the entropy of our system starts out being nonzero because of the
thermalization in the vertical variable. We are really interested in
whether or not $S$ increases from this initial offset, because that
change would be caused by growing horizontal disorder.
But we have seen that a system initially in a horizontal eigenstate
experiences no entropy increase for
any value of $b$ in the large $Q$ limit. This means that the system evolves
in a pure-state manner (subtracting off the initial vertically induced
offset), despite having its dynamics dominated by interactions
with the bath: environmental decoherence hits a maximum at an intermediate
value of $Q$, not in the limit $Q \to \infty$. For $b \neq 0$, this is in a
sense Pyrrhic victory over decoherence since it comes at the cost of
freezing. For $b = 0$ on the other hand,
the ensemble described by the reduced density matrix
evolves nontrivially
as if it were a single isolated particle undergoing unitary
evolution at an averaged frequency.

Whether freezing or synchronization results in the large $Q$ limit
depends on how the bath couples to the system. If the bath is side-sensing,
($b \neq 0$) then freezing happens;
if it is side-blind ($b = 0$), then synchronization occurs.
It has often been remarked that the QZE case is connected with
the measurement process.
Since the environment has a $b \neq 0$ coupling, it can ``tell the
difference'' between left and right sides of the well, and thus
``measure'' this observable (the Hermitian operator $\sigma_3$).
When $Q \to \infty$, the system-bath ``measurement'' interactions
happen repeatedly and quickly (constant monitoring),
forcing the system to remain in the
same $\sigma_3$ eigenstate (or more generally to remain in the
same statistical mixture of eigenstates). On the other hand, if
the coupling is of $b = 0$ form, then there is no sense
in which the environment probes whether the system is in the
left or right state, and therefore it cannot ``measure''
and thus freeze $\sigma_3$. The oscillations between left
and right are undamped, but the system states transform
between vertical $E_i$ levels so rapidly that they
effectively all oscillate at the same averaged frequency.

These are heuristic observations, as indicated by the scare quotes
around key phrases and words. It would be useful to make the
connection between $b$ and measurement more rigorous.
This should
be possible in light of the recent clarification of how
exactly measurement can be modeled in quantum theory.

The emerging paradigm connects the measurement process to
entanglement between system, apparatus and environment states \cite{zurek}. 
We will see that microscopic collisions between a system particle and a spin-$0$
boson from the bath can be interpreted as inducing this sort of
entanglement in the $b \neq 0$ case.\footnote{System-environment
entanglement should not be confused with the intra-system
entanglement studied in Sec.\ref{entangle}.}
So, the microscopic perspective will provide us with a simple explanation for
why the $b \neq 0$ and $b = 0$ cases are different, and indeed why the
former is connected with measurement. These considerations will not
capture the full complexity of the phenomena generated by the master equations
(synchronization in particular), but they will furnish some physical insight.

We first briefly summarize the emerging paradigm.
The world is divided into three parts: system, apparatus and environment.
Suppose that the system is initially in state
$|{\cal S}\rangle = \sum_n c_n(0) |n\rangle$,
where $|n\rangle$ is an eigenstate of $O$, the system
observable that the apparatus
can measure. Initially, we take the apparatus and environment to be in
states $|{\cal A}\rangle$
and $|{\cal E}\rangle$, respectievly. We also suppose that
initially there are no correlations between the three subsystems, so
the state of the world begins as simply $|{\cal S}\rangle
\otimes |{\cal A}\rangle \otimes |{\cal E}\rangle$. By time $t_1$, the
system and apparatus have unitarily evolved into the entangled state
$\sum_n c_n(t_1) |n\rangle \otimes |a_n\rangle$, so that a correlation
now exists between eigenstates of the system $|n\rangle$ and some basis states
$|a_n\rangle$ for the apparatus. In the typical picture, the environment at this
``pre-measurement''
stage remains uncorrelated with the system-plus-apparatus hybrid. 
The measurement is not complete at this stage because quantal 
correlations exist between the ``outcomes''. By time
$t_2$, however, we suppose that unitary evolution has also entangled
the environment with the 
apparatus, in a way such that each apparatus state $|n\rangle$ is linked to 
a distinct environmental state, $|e_n\rangle$. Now the world is in the fully
correlated state
\begin{equation}
\sum_n c_n(t_2) |n\rangle \otimes |a_n\rangle \otimes |e_n\rangle.
\label{fullycorrelated}
\end{equation}
The final step is to argue that the environment states $|e_n\rangle$
are by definition
inaccessible or uninteresting, so one should trace over them to obtain
the reduced density matrix for system-plus-apparatus. Doing this we
obtain
\begin{equation}
\rho_{\text{s}+\text{a}}(t_2) = \sum_n |c_n(t_2)|^2 |n\, ;\, a_n \rangle
\langle n\, ;\, a_n|,
\end{equation}
where we have taken the environment states to be orthonormal,
and we have defined $|n\, ;\, a_n\rangle \equiv |n\rangle \otimes |a_n\rangle$.
This reduced density matrix describes a mixture of states
which have {\it classically}
correlated system-apparatus variables. The measurement is complete,
with all {\it quantal} correlations between system and apparatus having
diffused into the environment, thus effectively becoming lost.

We will now connect this schematic to our case by examining a microscopic
energy-changing collision between a system particle and a spin-$0$ bath boson
$\phi$. We previously called the Hilbert space spanned by the horizontal
{\it and} vertical states as the ``system''. It is now useful to change perspective.
Let us redefine ``system'' to mean the Hilbert space
spanned by just the horizontal eigenstates (left or right), with the
``apparatus'' being spanned by the vertical states (single well energy
eigenstates labeled by $E$).
The environment is spanned by the energy states $|e\rangle$ of $\phi$.

The initial state is
\begin{equation}
\left[ c_1(0) |L\, ,\, E\rangle + c_2(0) |R\, ,\, E\rangle \right] \otimes |e\rangle,
\label{initialstate}
\end{equation}
the left and right eigenstates being
\begin{eqnarray}
|L\, ,\, E\rangle & = & \frac{ |E + \omega\rangle + |E - \omega\rangle }{\sqrt{2}},\nonumber\\
|R\, ,\, E\rangle & = & \frac{ |E + \omega\rangle - |E - \omega\rangle }{\sqrt{2}},
\end{eqnarray}
respectively, where $E \pm \omega$ label the nearly degenerate double-well states
collectively indexed by the vertical designation $E$. Free unitary evolution
takes the inital state in Eq.(\ref{initialstate}) to
\begin{equation}
e^{-i(E + e)t} \left[ c_1(t) |L\, ,\, E\rangle + c_2(t) |R\, ,\, E\rangle \right]
\otimes |e\rangle
\end{equation}
at time $t$, where $c_{1,2}(t) = c_{1,2}(0) \cos \omega t - i c_{2,1}(0) \sin \omega t$.
We will ignore the unimportant overall phase factor in the following.
Suppose a collision with a bath boson occurs at the time $t$ (we are assuming that the
interaction time is very short compared to the mean free path).
For simplicity, consider the extreme case $b = 1$ so that only $L$ states interact
with $\phi$, and work in one spatial dimension. The latter idealisation,
which is inessential, leads to the final energies being completely determined
by kinetic energy and momentum conservation from the initial energies $E$ and $e$.
Immediately after the collision, the state of the system is
\begin{equation}
c_1(t) |L\, ,\, E'\rangle \otimes |e'\rangle
+ c_2(t) |R\, ,\, E\rangle \otimes |e\rangle,
\label{beq1}
\end{equation}
where $E' \neq E$ and $e' \neq e$.\footnote{In addition, the oscillation frequency has changed from
$\omega$ to $\omega'$ for the left eigenstate.}
This is already in the fully entangled form of Eq.(\ref{fullycorrelated}); there
is no distinct pre-measurement step. We now
trace over the $\phi$ energy states to obtain the reduced density matrix
\begin{equation}
|c_1(t)|^2 |L\, ,\, E'\rangle \langle L\, ,\, E'| +
|c_2(t)|^2 |R\, ,\, E\rangle \langle R\, ,\, E|.
\end{equation}
In the sense made precise by the comparison of the above simple argument
with the general schema, the measurement of the horizontal variable is complete.

On the other hand, if $b = 0$, then Eq.(\ref{beq1}) is replaced by
\begin{equation}
\left[ c_1(t) |L\, ,\, E'\rangle
+ c_2(t) |R\, ,\, E'\rangle \right] \otimes |e'\rangle,
\label{beq2}
\end{equation}
whose reduced density matrix is that for the pure state
$c_1(t) |L\, ,\, E'\rangle  + c_2(t) |R\, ,\, E'\rangle$.
There is no entanglement with the environment.
The horizontal quantity remains unmeasured, even though
there has been an energy changing collision with a bath boson.

It is interesting that it makes sense to consider the vertical states as
spanning an apparatus which measures the horizontal states. We began by
considering a three-part division of the world (horizontal, vertical, bath)
for other reasons, but apparently this same partition provides a framework
for interpreting aspects of the $b \neq 0$ dynamics as a
canonical measurement process. Indeed, our master equation
incorporating both horizontal and vertical states then precisely specifies
the dynamics of what can be interpreted as a system-plus-apparatus hybrid.

We can sharpen this point. Suppose we put some particles into one of the energy
eigenstate of an isolated well. They oscillate between the left and right
states at a given frequency.
At time $t_{\text{bath}}$ (bath time) suddenly immerse it in a
$b = 1$, $Q \to \infty$ bath. If the particles' phases happen to place them
entirely in the
left state at this time, then the system quickly becomes thermalised in
energy and frozen in the left eigenstate. Now remove the system from the
bath. If a thermalised distribution of oscillating particles is
seen, then we know that the horizontal state was measured to be ``left''.
On the other hand, if the phases at $t_{\text{bath}}$ happen to place
all of the particles in the right eigenstate, then the system will remain
unchanged in energy but frozen in the right eigenstate. Upon removal
from the bath, we see a non-thermalised distribution and hence
we know the measurement yielded ``right''.

\section{Conclusion}
\label{conc}

We have analysed behaviors that can be exhibited by a quantum system that
is coupled to a thermal bath. The system states were taken to have both a horizontal
and a vertical structure. The horizontal states are discrete and the
system is generally in a superposition of them. Familiar examples
are the left and right sides of a symmetrical double well, the case
we focussed on, and neutrino flavor. The vertical states are labeled by the
energy eigenvalues. A ``dynamical superselection rule''
induced by decohering interactions with the bath quickly eliminates
superpositions in the vertical direction. However, the same system-bath
coupling induces system transitions between the vertical states. The
limit where the system-bath interaction rates were much larger than
the free oscillation rates of the system was our particular concern.

Two qualitatively different, but complementary, limiting
cases were analysed. The frozen or quantum Zeno case arises when the system-bath
interactions distinguish between the horizontal states (side-sensing). The synchronized or
motionally narrowed case arises, by contrast, when the system-bath
interactions are blind to the horizontal structure (side-blind). The latter can only
be observed when the vertical structure is present, whereas the former obtains
even when the system is vertically trivial. Two extensions were made:
to a double-well containing fermions, where the effect of Fermi statistics
was fully incorporated into the master equation; and to potentials with a
time-dependent bias. In the synchronization limit, the multi-fermion system
presents a behavior which gives the illusion that all of the fermions are
in the same state, evolving coherently. When there is a time-dependent bias,
the synchronization limit can lead to an increase in the efficiency of
transformation from one horizontal state into the other.

We then examined the time evolution of bipartite entanglement-of-formation
in the face of side-blind system-bath interactions. Entanglement
was observed to at first decrease with increasing system-bath coupling strength.
However, when the synchronization limit was reached, entanglement was found
to be preserved.

The connection between freezing/QZE on the one hand, and
synchronization/motional-narrowing on the other, and measurement was clarified.
By adopting the emerging ``system + apparatus + environment''
paradigm for quantal measurement, we were able to show that microscopic
energy-changing interactions between a system particle and a bath boson
can be naturally interpreted as measurement events in the horizontal
observable. The tripartite subdivision ``horizontal + vertical + bath''
was mapped on to ``system + apparatus + environment'' in the process.

In closing, we would like to note that the synchronization or motional narrowing
limit should be useful in applications such as quantum computation
which exploit quantal coherence and entanglement. The experimental or
engineering challenge would be to construct devices whose active
or horizontal system elements are deliberately placed in a noisy
environment, but one that couples to all horizontal states equally strongly.

\acknowledgements

The work of R.F.S. was supported, in part, by NSF grant PHY-9900544,
N.F.B. by the Commonwealth of Australia and the Australian Research Council, 
and R.R.V. by the Australian Research Council. 
We thank Y. Y. Y. Wong for helpful discussions.

\appendix

\section{Derivation of the master equations}

We provide a derivation of the master equations appropriate for the context of this paper.
We focus on the aspects that give rise to the placing of the horizontal matrices in Eq.(\ref{3})
and the detailed form of the right hand side 
of Eq.(\ref{11}). The methods are closely related to those given in
\cite{vankampen} but we employ no Markov approximation.
We introduce an interaction picture based on the above division of the Hamiltonian.
The
system-bath coupling in this picture is denoted by $H_I(t)$,  with vertical matrix elements,
\begin{equation}
 [ H_I(t)]_{j,k}=e^{i\lambda(E_j) t}\zeta e^{-i\lambda(E_k)
t}V^I_{j,k}e^{i(E_j-
E_k)t}.
\label{5}
\end{equation}
The leading terms in the system+bath density matrix, $\rho^{\rm
s+b}_{I}$, in this interaction picture are generated by the integral equation,

\begin{eqnarray}
\rho^{\rm s+b}_I (t) &=& \rho^{\rm s+b}(t=0) \nonumber\\
&-&\int_0^t dt_1\int_0^{t_1}dt_2\Bigr[
[\rho^{\rm s+b}_I(t_2),H_I(t_2) ],H_I(t_1)\Bigr ].
\label{6}
\end{eqnarray}

In the perturbation expansion of Eq.(\ref{6}) we find a piece with an additional power of the quantities
$c_{j,k}^2 [E t, E \lambda^{-1}]$ in each higher order. This provides our definition of leading terms
and the explanation for how a tiny $c_{j,k}^2$ can lead to appreciable effects over long periods of
time. These terms are generated by the terms in which each successive new pair of $H_I$'s in the
iteration solution puts the state back to the same vertical (and bath) state. These observations
provide the basis for having omitted, in writing 
Eq.(\ref{6}), all odd terms in $H_I$, and they also
dictate the form of the density matrix to be used in solving the equation,
\begin{equation}
\rho^{\rm s+b}_I=\rho^{\rm bath}\sum_j \rho_I(E_j,t) |j \rangle \langle j|.
\label{7}
\end{equation}
where,
\begin{equation}
\rho^{\rm bath}=\Bigr[\prod_i(1-e^{-\omega_i/T})\Bigr]\exp\Bigr [-(\sum_n a_n
^\dagger a_n \omega_n)/T \Bigr ],
\label{8}
\end{equation}
and where,
\begin{equation}
\rho_I(E_j,t)=e^{i\lambda (E_j)t}\rho(E_j,t)e^{-i\lambda (E_j)t}.
\label{9}
\end{equation}
We substitute  Eq.(\ref{7}) into Eq.(\ref{6}), take the bath trace, and use leading
order
expressions for each term in the double commutator, obtaining, for example,
\begin{eqnarray}
&{\rm Tr}&_{\rm bath}\Bigr[\langle j|H_I(t_1)\rho^{\rm s+b}_I
(t_2)H_I(t_2)|j\rangle \Bigr] =\delta(t_1-t_2)  \nonumber\\
&\times& \sum _k \Gamma(E_k,E_j)
e^{i\lambda(E_j) t_1}\zeta e^{-i\lambda(E_k)
t_1} \nonumber\\
&\times& \rho_I(E_k,t_1)e^{i\lambda(E_k)
t_1}\zeta e^{-i\lambda(E_j) t_1}.
\label{10}
\end{eqnarray}
Similarly, we have
\begin{eqnarray}
&{\rm Tr}&_{\rm bath}\Bigr[\langle j|H_I(t_1)H_I(t_2)\rho_I^{\rm
s+b}(t_2)|j \rangle \Bigr]=\delta(t_1-t_2) \nonumber\\
&\times& e^{i\lambda(E_j)
t_1}\zeta^2e^{-i\lambda(E_j
)t_1}\sum _k \Gamma(E_j,E_k) \rho_I(E_j, t_1).
\label{12}
\end{eqnarray}
The key to separation of the leading terms in the above is performing the sum
over the
modes of the scalar field first. After making the transition to the continuum,
we make replacements of the form,
\begin{eqnarray}
&\int_{\omega_{min}}^\infty & d \omega f(\omega) e^{i(\pm E_j \mp E_k \pm
\omega)(t_1-t_2)} \nonumber\\
&\rightarrow&
2\pi
f[(E_k-E_j)]\delta(t_1-t_2)\theta(E_k-E_j).
\label{13}
\end{eqnarray}
where neither the introduced infrared 
cut-off, $\omega_{min}$ nor the residual terms contribute to leading
order, as defined above.

Using  Eqs.(\ref{10}), (\ref{12}) and their counterparts for the two other
orderings in
Eq.({6}), doing the $t_2$ integral (with the delta function symmetrically smeared, as can be
justified by more accurate integrations) and differentiating  with respect to $t$ gives,
\begin{eqnarray}
&{d\over dt}&\rho_I(E_j,t)=\sum_{E_k}e^{i\lambda(E_j) t}\zeta
e^{-i\lambda(E_k) t}
\rho_I(E_k,t) \nonumber\\
&\times& e^{i\lambda(E_k) t}\zeta e^{-i\lambda(E_j) t}
\Gamma(E_k,E_j)
\nonumber\\
&-&{1\over 2}\Bigr(e^{i\lambda(E_j) t}\zeta^2e^{-i\lambda(E_j )t}
\rho_I(E_j,t) \nonumber\\
&+&\rho_I(E_j,t) e^{i\lambda(E_j)
t}\zeta^2e^{-i\lambda(E_j )t}
\Bigr)
\sum_{E_k}\Gamma(E_j,E_k).
\label{14}
\end{eqnarray}
Using Eq.(\ref{9}) we now
regain Eq.(\ref{3}).


\begin{thebibliography}{99}


\bibitem{leggett} A. J. Leggett, S. Chakravarty, A.T. Dorsey, M.P.A. Fisher,
A. Garg, and  W. Zwerger, Rev. Mod. Phys. {\bf 59}, 1 (1987) and references therein.

\bibitem{nielsen}
For an overview see
M. A. Nielsen and I. L. Chuang, {\it Quantum computation and quantum information},
Cambridge University Press (2000). 

\bibitem{reviews}
For recent reviews see
M. Prakash, J. M. Lattimer, R. F. Sawyer and R. R. Volkas, 
Ann. Rev. Nucl. Part. Sci. (in press, 2001), astro-ph/0103095;
P. Di Bari, R. Foot, R. R. Volkas and Y. Y. Y. Wong, Astropart. Phys. {\bf 15}, 391 (2001).

\bibitem{zurek}
See, for example, W. H. Zurek, Physics Today {\bf 44} (October) 36 , (1991);
quant-ph/0105127; 
J. P. Paz and W. H. Zurek, Proceeding of the 72nd Les Houches Summer 
School on ``Coherent Matter Waves'', July-August 1999, quant-ph/0010011;
Roland Omn\`{e}s, {\it Understanding quantum mechanics}, Princeton University Press (1999);
D. Giulini et al., {\it Decoherence and the appearance of a classical world 
in quantum theory}, Springer (1996).

\bibitem{zeno}
B. Misra and E. C. G. Sudarshan, J. Math. Phys. {\bf 18}, 756 (1977).

\bibitem{harrisandstodolsky}
 R. A. Harris and L. Stodolsky,
 Phys. Lett. {\bf 116B}, 464 (1982);

\bibitem{silbey} R. A. Harris and R. Silbey, J. Chem. Phys. {\bf 78}, 7330 (1993)

\bibitem{silvestrini}
P. Silvestrini and L. Stodolsky, Phys. Lett. {\bf A280}, 17 (2001);
P. Silvestrini and L. Stodolsky, cond-mat/0010129.

\bibitem{earlyuniverse}
L. Stodolsky, Phys. Rev. {\bf D36}, 2273 (1987).
See also
A. Dolgov, Sov. J. Nucl. Phys. {\bf 33}, 700 (1981);
K. Enqvist, K. Kainulainen and J. Maalampi,  Nucl. Phys. {\bf B 349}, 754 (1991).


\bibitem{bell} N. F. Bell, R. F. Sawyer and R. R. Volkas, Phys. Lett. {\bf B500}, 16
(2001).

\bibitem{motionnarrow}
A. Abragam, {\it The principles of nuclear magnetism}, Oxford (1961), ch. 10.
See also Eberly et al., Phys. Rev. A30, 2381 (1984) for a ``motion narrowing'' effect 
in optics.

\bibitem{refrigerator}
W. H. Zurek, Phys. Rev. Lett. {\bf 53}, 391 (1984);
W. Y. Hwang et al., Phys. Rev. {\bf A 62}, 062305 (2000), and references
therein.



\bibitem{MandT} B. H. J. McKellar and M. J. Thomson, Phys. Rev.  {\bf  D49}, 2710 (1994).


\bibitem{raffelt} 
G. Raffelt, G. Sigl, and L. Stodolsky, Phys. Rev. Lett. {\bf 70}, 2363 (1993).  
See also G. Raffelt and G. Sigl, Astropart. Phys. {\bf 1}, 165 (1993).


\bibitem{msw}
L. Wolfenstein, Phys. Rev. {\bf D17}, 2369 (1978); {\it ibid.} {\bf D20},
2634 (1979);
S. P. Mikheyev and A. Yu. Smirnov, Nuovo Cimento {\bf C9}, 17 (1986).
For a review see e.g.
T. K. Kuo and J. Pantaleone, Rev. Mod. Phys. {\bf 61}, 937 (1989).



\bibitem{wootters}
S. Hill and W. K. Wootters, Phys. Rev. Lett. {\bf 78}, 5022 (1997);
W. K. Wootters, Phys. Rev. Lett. {\bf 80}, 2245 (1998).


\bibitem{vankampen}N. G. van Kampen {\it Stochastic Processes in Physics and Chemistry}
North Holland (Amsterdam, 1992) (revised and enlarged edition), ch. XVII;
D. F. Walls and G. C. Milburn, {\it Quantum Optics}, Springer (Berlin, 1994), ch. 6.

















\end{thebibliography}
\end{document}